\documentclass[aps,prd,nofootinbib,onecolumn,notitlepage]{revtex4-1}
\usepackage[utf8]{inputenc}
\usepackage{amsmath}
\usepackage{amsfonts}
\usepackage{amssymb}
\usepackage{color}
\usepackage[unicode=true,pdfusetitle,
 bookmarks=true,bookmarksnumbered=false,bookmarksopen=false,
 breaklinks=false,backref=false,colorlinks=false]
 {hyperref}

\newcommand{\cob}{\color{blue}}

\renewcommand\[{\begin{equation}}
\renewcommand\]{\end{equation}}
\newcommand{\al}{\alpha}
\newcommand{\bt}{\beta}
\newcommand{\ga}{\gamma}

\newcommand{\n}{\nabla}
\newcommand{\ba}{\begin{eqnarray}}
\newcommand{\ea}{\end{eqnarray}}

\begin{document}
\title{Towards resolution of anisotropic cosmological singularity in infinite derivative gravity}
\author{Alexey S. Koshelev$^{1,2,3,4}$,~Jo\~ao Marto$^{1,2}$,~Anupam Mazumdar$^{5,6}$}
\affiliation{$^{1}$Departamento de F\'isica, Universidade da Beira Interior, Rua Marquês D'Ávila e Bolama, 6201-001 Covilhã, Portugal.}
\affiliation{$^{2}$Centro de Matemática e Aplicações da Universidade da Beira Interior
(CMA-UBI), Rua Marquês D'Ávila e Bolama, 6201-001 Covilhã, Portugal.}
\affiliation{$^{3}$Theoretische Natuurkunde, Vrije Universiteit Brussel.} 
\affiliation{$^{4}$The International Solvay Institutes, Pleinlaan 2, B-1050, Brussels, Belgium.}
\affiliation{$^{5}$Van Swinderen Institute, University of Groningen, 9747 AG, Groningen, The Netherlands}
\affiliation{$^{6}$Kapteyn Astronomical Institute, University of Groningen, 9700 AV Groningen, The Netherlands.}
\date{\today}
\begin{abstract}
In this paper, we will show that the equations of motion of the quadratic in curvature, {\it ghost free}, infinite derivative theory of gravity will not 
{\it permit} an anisotropic collapse of a homogeneous Universe for a Kasner-type vacuum solution. 
\end{abstract}
 \vskip -1.0 truecm

\maketitle

\section{Introduction}
\label{Introduction }

Einstein's theory of general relativity (GR) provides non-trivial vacuum solutions with cosmological and blackhole type singularities~\cite{Hawking:1973uf}. 
In the context of a blackhole, within GR, the cosmic censorship forbids naked singularity~\cite{Penrose,Penrose:1964wq,Wald:1997wa}, but in the cosmological context the singularity is not covered by any horizon, and leads to an incomplete null/time like-geodesics and subsequently breakdown of classical and quantum initial conditions, see~\cite{Geroch:1968ut}, and \cite{Wald:1984rg}. In particular, the Kasner metric~\cite{Kasner:1921zz}, which is well known to yield a Belinsky-Khalatnikov-Lifshitz (BKL) type singularity~\cite{Belinsky:1982pk,Belinsky:1970ew}, where the metric becomes singular as $t\rightarrow 0$. Indeed, it is a wishful thinking to resolve both cosmological and blackhole type singularities, at a classical and at a quantum level. 

Recent work by Biswas, Gerwick, Koivisto and Mazumdar (BGKM)~\cite{Biswas:2011ar} have shown that the  quadratic curvature, {\it ghost free} infinite derivative theory of gravity in $4$ spacetime dimensions can avoid both cosmological and blackhole type singularities at the linearized level around the 
Minkowski background~\cite{Biswas:2011ar}~\footnote{See previous to this work other relevant references~\cite{Tseytlin:1995uq,Siegel:2003vt,Biswas:2005qr,Biswas:2006bs}, 
where the authors have argued absence of singularity in infinite derivative gravity motivated from the string theory, however, the full quadratic curvature action including the Weyl term with two gravitational metric potentials were first presented in~\cite{Biswas:2011ar}. The other constant curvature backgrounds have also been considered, such as de Sitter and anti de Sitter backgrounds~\cite{Biswas:2016etb,Biswas:2016egy}.},
while the cosmological singularity can be resolved even at the full non-linear level, but the previous works have sought for a bouncing solution, within the infinite derivative Ricci scalar, with one extra scalar propagating degree of freedom, besides the graviton, which are both ghost-free~\cite{Biswas:2005qr,Biswas:2006bs,Koivisto,Koshelev:2012qn,Conroy:2014dja}. In Ref.~\cite{Biswas:2011ar}, the homogeneous and isotropic bouncing solution was constructed in a vacuum, with the full {\it ghost free} infinite derivative quadratic curvature action at the linear level, which is not otherwise possible within GR.

At a linear level (around asymptotically Minkowski background), resolution of blackhole singularities has also been studied both in the context of static~\cite{Biswas:2011ar,Biswas:2013cha,Frolov-1,Edholm:2016hbt,Frolov,Koshelev:2017bxd,Buoninfante:2018xiw}, for extended objects \cite{Boos:2018bxf}, and in a 
rotating blackhole case~\cite{Cornell:2017irh} by various groups. Furthermore, lack of dynamical formation of singularity at the linear level has also been studied by Frolov and his collaborators~\cite{Frolov:2015bia,Frolov:2015usa}. Furthermore, the gravitational force quadratically vanishes towards the center, in the linear regime, and it has been shown that a compact astrophysical object can be formed, which is devoid of curvature singularity and the event horizon all together~\cite{Koshelev:2017bxd}.  Since, all the interactions are {\it purely} derivative in nature, the gravitational {\it form factors} give rise to non-local interactions for the BGKM action~\cite{Tomboulis,Tomboulis:2015,Modesto,Talaganis:2014ida,Biswas:2014yia}. 

Recently an interesting progress has been made to show that the BGKM gravity with full {\it non-linear}  equations of motion, given by~\cite{Biswas:2013cha}, will not permit singular solution of type $1/r^{\alpha}$, where $\alpha >0$, in the static asymptotically Minkowoski background in $4$ dimensions~\cite{Koshelev:2018hpt}. The results of the paper exploit the importance of the quadratic curvature infinite derivative Weyl contribution, and argue that the higher covariant derivative terms {\it do not } satisfy the vacuum condition, order-by-order. The analysis also sheds light on why {\it just} the local quadratic curvature action, including Weyl, is {\it insufficient} to resolve the singular solution of the type $1/r^{\alpha}$~\cite{Koshelev:2018hpt}.

The aim of this paper will be to show that the full non-linear equations of motion of the BGKM gravity will not permit Kasner solution~\cite{Kasner:1921zz}, in a vacuum, in $4$ dimensions. We will  now briefly recall the BGKM gravity and its equations of motion and study the viability of a Kasner solution, with the Ricci scalar and the Ricci tensor to be vanishing in the vacuum.


\section{Introduction to infinite derivative gravity}

The most general diffeomorphism and parity invariant quadratic curvature action, but free from torsion, has been derived in Ref.\cite{Biswas:2011ar,Biswas:2013cha}, given by~
\footnote{The action in Ref.\cite{Biswas:2011ar} was first written in terms of the Riemann tensor, here we will work in terms of the Weyl tensor, which can be rewritten in terms of the Riemann tensor as: \[
\label{weyl}
W_{\;\alpha\nu\beta}^{\mu}=R_{\;\alpha\nu\beta}^{\mu}-\frac{1}{2}(\delta_{\nu}^{
\mu}R_{\alpha\beta}-\delta_{\beta}^{\mu}R_{\alpha\nu}+R_{\nu}^{\mu}g_{
\alpha\beta}-R_{\beta}^{\mu}g_{\alpha\nu})+\frac{R}{6}(\delta_{\nu}^{\mu}g_{
\alpha\beta}-\delta_{\beta}^{\mu}g_{\alpha\nu})
\]
}
\begin{equation}\label{action}
S=\frac{1}{ {16\pi G} }\int d^{4}x\sqrt{-g}\left(R+\alpha_c\left[R{\cal
F}_{1}(\Box_s)R+R^{\mu\nu}{\cal
F}_{2}(\Box_s)R_{\mu\nu}+W^{\mu\nu\lambda\sigma}{\cal
F}_{3}(\Box_s)W_{\mu\nu\lambda\sigma}\right]\right)\,,
\end{equation}
where $G=1/M_p^2$ is the Newton's gravitational constant, and $\alpha_c \sim 1/M_s^2$ is a dimensionful coupling, where $\Box_s\equiv \Box/M_s^2$. The $M_s$ signifies the scale of non-local interactions in gravity. In the limit $M_s\rightarrow \infty$, the action reduces to the 
Einstein-Hilbert one. The d'Alembertian operator is defined as: $\Box=g^{\mu\nu}\nabla_{\mu}\nabla_{\nu}$, where $\mu\,,\nu=0,1,2,3$, and we work with mostly positive metric convention $(-,+,+,+)$. The ${\cal F}_{i}$'s are three gravitational {\it form-factors}, which are defined by:
\begin{equation}
{\cal F}_{i}(\Box_s) =\sum_{n\geq 0} c_{i, n}\Box_s^{n}\,.
\end{equation}
The coefficients $c_{i, n}$ are constrained by the fact that the graviton propagator for the full action {\it only} contains the transverse and trace-less graviton degrees of freedom, and no extra dynamical degrees of freedom~\cite{Biswas:2013kla}. Indeed, such an action introduces non-local gravitational interaction, which is indeed helpful to ameliorate the quantum aspects of the theory in the ultraviolet~\cite{Tomboulis,Tomboulis:2015,Modesto,Talaganis:2014ida}.

The complete equations of motion derived from the above action Eq.(\ref{action}) is given by \cite{Biswas:2013cha},
\begin{align}
P^{\alpha\beta}=& -\frac{G^{\alpha\beta}}{{ 8\pi G}} +\frac{\alpha_c}{{ 8\pi G}} \biggl( 4G^{\alpha\beta}{\cal
F}_{1}(\Box_s)R+g^{\alpha\beta}R{\cal
F}_1(\Box_s)R-4\left(\triangledown^{\alpha}\nabla^{\beta}-g^{\alpha\beta}
\Box_s\right){\cal F}_{1}(\Box_s)R
\nonumber\\&
-2\Omega_{1}^{\alpha\beta}+g^{\alpha\beta}(\Omega_{1\sigma}^{\;\sigma}+\bar{
\Omega}_{1}) +4R_{\mu}^{\alpha}{\cal F}_2(\Box_s)R^{\mu\beta}
\nonumber\\&
-g^{\alpha\beta}R_{\nu}^{\mu}{\cal
F}_{2}(\Box_s)R_{\mu}^{\nu}-4\triangledown_{\mu}\triangledown^{\beta}({\cal
F}_{2}(\Box_s)R^{\mu\alpha})
+2\Box_s({\cal
F}_{2}(\Box_s)R^{\alpha\beta})
\nonumber\\&
+2g^{\alpha\beta}\triangledown_{\mu}\triangledown_{
\nu}({\cal F}_{2}(\Box_s)R^{\mu\nu})
-2\Omega_{2}^{\alpha\beta}+g^{\alpha\beta}(\Omega_{2\sigma}^{\;\sigma}+\bar{
\Omega}_{2}) -4\Delta_{2}^{\alpha\beta}
\nonumber\\&
-g^{\alpha\beta}W^{\mu\nu\lambda\sigma}{\cal
F}_{3}(\Box_s)W_{\mu\nu\lambda\sigma}+4W_{\;\mu\nu\sigma}^{\alpha}{\cal {\cal
F}}_{3}(\Box_s)W^{\beta\mu\nu\sigma}
\nonumber\\&
-4(R_{\mu\nu}+2\triangledown_{\mu}
\triangledown_{\nu})({\cal {\cal F}}_{3}(\Box_s)W^{\beta\mu\nu\alpha})
-2\Omega_{3}^{\alpha\beta}+g^{\alpha\beta}(\Omega_{3\gamma}^{\;\gamma}+\bar{
\Omega}_{3}) -8\Delta_{3}^{\alpha\beta} \biggr)
\nonumber\\
=& -T^{\al\bt}\,,
\label{EOM}
\end{align}
where $T^{\al\bt}$ is the stress energy tensor for the matter components, where we have defined the following symmetric tensors, for detailed derivation, see~\cite{Biswas:2013cha}:
\begin{align}\label{details}
\Omega_{1}^{\alpha\beta}= & \sum_{n=1}^{\infty}f_{1_{n}}\sum_{l=0}^{n-1}\nabla^{
\alpha}R^{(l)}\nabla^{\beta}R^{(n-l-1)},\quad\bar{\Omega}_{1}=\sum_{n=1}^{\infty
}f_{1_{n}}\sum_{l=0}^{n-1}R^{(l)}R^{(n-l)},
\\
\Omega_{2}^{\alpha\beta}= & \sum_{n=1}^{\infty}f_{2_{n}}\sum_{l=0}^{n-1}R_{\nu}^{
\mu;\alpha(l)}R_{\mu}^{\nu;\beta(n-l-1)},\quad\bar{\Omega}_{2}=\sum_{n=1}^{
\infty}f_{2_{n}}\sum_{l=0}^{n-1}R_{\nu}^{\mu(l)}R_{\mu}^{\nu(n-l)}\,,
\\
\Delta_{2}^{\alpha\beta}= & \sum_{n=1}^{\infty}f_{2_{n}}\sum_{l=0}^{n-1}
[R_{
\sigma}^{\nu(l)}R^{(\beta\sigma;\alpha)(n-l-1)}-R_{\;\sigma}^{\nu;\alpha(l)
}R^{
\beta\sigma(n-l-1)}]_{;\nu}\,,
\\
\Omega_{3}^{\alpha\beta}= & \sum_{n=1}^{\infty}f_{3_{n}}\sum_{l=0}^{n-1}W_{
\: \: \nu\lambda\sigma}^{\mu;\alpha(l)}W_{\mu}^{\;\nu\lambda\sigma;\beta(n-l-1)},
\quad\bar{\Omega}_{3}=\sum_{n=1}^{\infty}f_{3_{n}}\sum_{l=0}^{n-1}W_{
\: \: \nu\lambda\sigma}^{\mu(l)}W_{\mu}^{\;\nu\lambda\sigma(n-l)}\,,
\\
\Delta_{3}^{\alpha\beta}= & \sum_{n=1}^{\infty}f_{3_{n}}\sum_{l=0}^{n-1}
[W_{\quad\sigma\mu}^{\lambda\nu(l)}W_{\lambda}^{\;\beta\sigma\mu;\alpha(n-l-1)}
-W_{\quad\sigma\mu}^{\lambda\nu\;\;;\alpha(l)}W_{\lambda}^{
\: \beta\sigma\mu(n-l-1)}
]_{;\nu}\,.\label{details-1}
\end{align}
The trace equation is rather simple, which can be written as~\cite{Biswas:2013cha}:
\begin{align}
P = &\frac{R}{{ 8\pi G}}+ \frac{\alpha_c}{{8\pi G}} \biggl( 12\Box_s{\cal F}_{1}(\Box_s)R+2\Box_s({\cal
F}_{2}(\Box_s)R)+4\triangledown_{\mu}\triangledown_{\nu}({\cal
F}_{2}(\Box_s)R^{\mu\nu})
\nonumber\\ &
+2(\Omega_{1\sigma}^{\;\sigma}+2\bar{\Omega}_{1})+2(\Omega_{2\sigma}^{\;\sigma}
+2\bar{\Omega}_{2})+2(\Omega_{3\sigma}^{\;\sigma}+2\bar{\Omega}_{3}
)-4\Delta_{2\sigma}^{\;\sigma}-8\Delta_{3\sigma}^{\;\sigma} \biggr)
\nonumber\\
= & -T\equiv
-g_{\al\bt}T^{\al\bt}\,.
\label{trace}
\end{align}
%

\section{Towards non-singular homogeneous and anisotropic metric}

Now, in order to show that the Kasner solution does not satisfy the equations of motion for the infinite derivative gravity, 
we will first assume that the above action, Eq.(\ref{action}), along with the equations of motion Eq.(\ref{EOM}), 
allows at least the vacuum solution, which is critical for the Kasner-type metric, if it had to be promoted as a solution, like in the case of GR~\footnote{We will be able
to relax this condition, see the discussion below.}
\begin{equation}\label{vacuum}
R=0\,,~~~~~~R_{\mu\nu}=0\,.
\end{equation}
The Kasner metric is given by~\cite{Kasner:1921zz}
\begin{equation}
ds^{2}=dt^{2}+ t^{2p_1} dx^{2} + t^{2p_2} dy^{2} + t^{2p_3} dz^{2} \;,
\label{metric-1}
\end{equation}
where the parameters $p_1 ,\: p_2 , \: p_3$ are constrained as,
\begin{equation}
p_1 + p_2 + p_3 =1\,, \quad \text{and} \quad p_1^2 + p_2^2 + p_3^2 = 1 \; .
\end{equation}
These two conditions can be expressed by the Khalatnikov-Lifshitz parameter, $u$, by (see Ref.\cite{Belinsky:1970ew}), 
\begin{equation}
p_1 = - \dfrac{u}{1+u+u^2}\; , \quad\quad p_2 =  \dfrac{1+u}{1+u+u^2}\; , \quad\quad p_3 = \dfrac{u(u+1)}{1+u+u^2} \; .
\label{Kh-Lifshitz}
\end{equation}
For the range $u\geq1$, the parameter $u$ covers all possible real-valued parameters ($p_1 ,\: p_2 , \: p_3$), 
and since Eq.~(\ref{Kh-Lifshitz}) possess the following symmetries,
\begin{equation}
p_1 (u) =  p_1 \left(\dfrac{1}{u}\right)\,,\quad \quad p_2 (u) =  p_3 \left(\dfrac{1}{u}\right)\,, \quad \quad p_3 (u) =  p_2 \left(\dfrac{1}{u}\right)\,,
\label{u-sym}
\end{equation}
the region $u<1$ can be mapped onto the region $u\geq1$. 

In addition, when $u=0$, the Kasner metric Eq.(\ref{metric-1}) yields, $p_2=1,~p_1=p_3=0$, and the Riemann tensor vanishes. In this case,  
using a suitable coordinate transformation, it is possible to obtain the Minkowski metric. When the parameter $u \rightarrow + \infty$,
we have, again, $p_3=1$, while $p_1=p_2 =0$, therefore, the metric Eq.~(\ref{metric-1}) recovers the Minkowski limit.  Also note that $u=-1$ replicates the case for $u=0$ with vanishing Riemann tensor. These  statements
can be summarized succinctly by the Riemann tensor for the Kasner metric, which is given by:
\begin{equation}
R_{\mu \nu \lambda \sigma}(u) \backsim \dfrac{u(u+1)}{(1+u+u^2)^2}\, . 
\end{equation}
Since the Ricci tensor and the Ricci scalar are both zero the Weyl tensor coincides with the Riemann tensor.

It is known that in GR the Kasner spacetime has a singularity when $t\rightarrow 0$. However,
the process by which this singularity is approached is peculiar and involves a complex 
oscillatory behavior, as shown in Refs.\cite{Belinsky:1970ew,Belinsky:1982pk}. When $t$ decreases, a succession of Kasner epochs take place 
based on a periodic (or chaotic) simultaneous change of sign in the Kasner constants $p_1$ and $p_2$. Consequently,
the volume of a Universe described by a Kasner metric decreases approximately as $\thicksim t$, 
with two spatial directions~\footnote{While the third direction just contracts, until eventually this third direction switches its role with one of the first two. 
This switching happens as a consequence of the symmetries (\ref{u-sym}).} oscillating between contraction and expansion, and presenting bounces.
Each Kasner era corresponds to an expanding/contracting phase between successive bounces. Another important aspect,
in this scenario, is that an infinite sequence of Kasner eras take place when $t\rightarrow 0$. 
This infinite sequence of eras ($n$) can be properly labelled by a decreasing rule for the parameter $u_n$,
\begin{equation}
u_{n+1} = u_n - 1 \:  (\text{if}\:\:  u_n \geqslant 2) \quad u_{n+1} = \dfrac{1}{u_n - 1} \:  (\text{if}\:\:  1 \leqslant u_n< 2)\,.
\end{equation}
Let us now concentrate on the full equations of motion Eq.~(\ref{EOM}), and let us assume that there is a vacuum configuration, $P^{\alpha \beta}=0$,
with Eq.(\ref{vacuum}). In fact, if we are keen on understanding the Kasner solution at $t\rightarrow 0$, or in the context of BGKM gravity, $ t < 1/M_s$,
it is suffice to assume $R\sim C_1$ and $R_{\mu\nu} \sim C_2$, where $C_1,~C_2$ are constants for $t< 1/M_s$. This is due to the fact that for $t< 1/M_s$, we are probing the 
UV aspects of gravity, where the infinite derivatives play the major role compared to the Einstein-Hilbert part of the action. Indeed, one may neglect the contribution from $G_{\alpha\beta}$ from Eq.(\ref{EOM}), and $R\sim C_1$ and $R_{\mu\nu} \sim C_2$ would suffice to concentrate on the Weyl component {\it alone}, in which case we are left with the following 
terms in the full equations of motion:
 \begin{align}
 P^{\alpha\beta} = 0 =  P^{\alpha\beta}_3=& \frac{\alpha_c}{{ 8\pi G}} \biggl( -g^{\alpha\beta}W^{\mu\nu\lambda\sigma}{\cal
F}_{3}(\Box_s)W_{\mu\nu\lambda\sigma}+4W_{\;\mu\nu\sigma}^{\alpha}{\cal {\cal
F}}_{3}(\Box_s)W^{\beta\mu\nu\sigma}
\nonumber\\&
-8\triangledown_{\mu}
\triangledown_{\nu}{\cal {\cal F}}_{3}(\Box_s)W^{\beta\mu\nu\alpha}
-2\Omega_{3}^{\alpha\beta}+g^{\alpha\beta}(\Omega_{3\gamma}^{\;\gamma}+\bar{
\Omega}_{3}) -8\Delta_{3}^{\alpha\beta} \biggr) \; .
\label{P3}
 \end{align}
The aim is to show that in the BGKM gravity,  the homogeneous and anisotropic collapse of the metric can be avoided by not allowing the existence of a 
Kasner metric, i.e., Eq.(\ref{metric-1}).  As a necessary condition (but not sufficient) for the 
Kasner metric to satisfy the equations of motion, i.e., Eq.(\ref{EOM}), both sides of the equation must vanish identically. The failure to do so will imply that the 
Kasner metric {\it cannot} be a vacuum solution for the BGKM gravity.\\

Let us summarize some important observations:

\begin{enumerate}
\item ${\cal F}_{i}(\Box_s)$ contain an infinite series of $\Box_s$. Indeed, the coefficients are not arbitrary as we had discussed briefly, they are pre-determined by the choice of {\it ghost-free} condition, i.e., the propagator of the BGKM gravity is suppressed by {\it exponential of an entire function}, as shown in Refs.~\cite{Biswas:2011ar}.

\item The Bianchi identity holds for each and ever order in $\Box_s$, see discussion in Ref.\cite{Koshelev:2018hpt}. The order is  intrinsically parametrized by the 
power $n$ of $\Box_s^n$.

\item The right hand side of Eq.~(\ref{P3}) should vanish at each and every order in $\Box_s$. 
The reason is that every $\Box_s$ would give rise to an extra factor of $1/t^2$. We are also assuming
that the parameters ($p_1 ,\: p_2 , \: p_3$) do not give rise to the trivial Minkowski solution, or  $p_2=1,~p_1=p_3=0$, as we had already discussed. 
Note that, for a given power $n$ of the d'Alembertian operator, the corresponding contribution from the Weyl part does not vanish automatically, nor the entire sum will vanish. The latter has 
a slim possibility, but would require extreme fine tuning, given the way the series progresses as $\propto 1/t^{\beta}$, where $\beta$ is an even number. The only chance to get rid of such a term is to adjust the coefficient $c_{3,n}$ to be equal 0, which is not the case for the BGKM action. 
\end{enumerate}

Given all these salient features, we will study what happens at each and every order in $\Box_s$. If the Kasner solution has to be admitted, then each and every order in $\Box_s$, the right hand side of Eq.(\ref{P3}) must vanish.
We can show that
the full computation for the right hand side of Eq.(\ref{P3}) would yield:
\begin{equation}
	P^{\alpha\beta}_{3}=\sum_{n\geq 1}f_{3n} \frac{w^{\alpha\beta}_n}{t^{4+2n}}\,,
\end{equation}
where coefficients $w^{\alpha\beta}_n$ are constants with respect to time $t$, and depend only on parameter $u$. We can also show that the contribution coming from the part without any $\Box_s$, i.e. the local contribution from the  Weyl squared gravity,  yields identically zero in 4 dimensions, 
thanks to the existence of the Gauss-Bonnet topological invariance, which means that the Kasner is a good solution for the local quadratic curvature gravity. In the appendix, we have collected the details of an explicit computation of $w^{\alpha\beta}_1$ and  $w^{\alpha\beta}_2$.

To summarize, in this paper, we have presented strong arguments that the homogeneous, anisotropic collapse of a Kasner metric in a vacuum cannot be a 
solution of the full infinite derivative gravity given by Eq.(\ref{action}).
A very similar conclusion we have reached for the static Schwarszchild-type  metric in Ref.\cite{Koshelev:2018hpt}. The presence of infinite 
derivatives indeed ameliorate the cosmological singularity. Indeed, how the time dependent metric will behave near $t\leq 1/M_s$ is still
an open question, but it would be extremely unlikely that the solution would yield a cosmological singularity in $4$ dimensions. 
The central theme of the resolution appears to be the Weyl squared contribution with infinite covariant derivatives in the equations of motion for the BGKM 
gravity. 

\section{Appendix}

\subsection{Non-vanishing contributions from the non-local Weyl term}

First, we explicitly compute one  $\Box_s$  contribution to Eq.(\ref{P3}), assuming the parameter redefinitions from Eq.(\ref{Kh-Lifshitz}) for the metric Eq.(\ref{metric-1}). Let us define 
$$P_3^{\alpha \beta }(\Box_s)= \frac{\alpha_c}{{ 8\pi G}} \sum_{i=1}^{6} F_i^{\alpha \beta }.$$

\begin{enumerate}
\item For the first term, $F_1^{\alpha \beta }=-g^{\alpha\beta}W^{\mu\nu\lambda\sigma}(f_{30}+f_{31}\Box_s)W_{\mu\nu\lambda\sigma}$, the calculation yields
\begin{equation}
F_1^{\alpha \beta }= g^{\alpha \beta}\frac{16}{M_{s}^2 \: t^6} \left(
\begin{array}{cccc}
 a_{00}(u) & 0 & 0 & 0 \\
 0 & a_{11}(u) & 0 & 0 \\
 0 & 0 & a_{22}(u) & 0 \\
 0 & 0 & 0 & a_{33}(u) \\
\end{array}
\right)\:,
\end{equation}
with the dimensionless matrix elements $a_{ii}$ defined as
\begin{equation}
a_{00} = a_{11} = a_{22} = a_{33} =-\frac{u^2 (u+1)^2 \left(f_{30} \: t^2 \: M_s^2 \left(u^2+u+1\right)^3  
- 9 f_{31} u^2 (u+1)^2 \right)}{ \left(u^2+u+1\right)^6 }\,. \nonumber
\end{equation}

\item The second term,
$F_2^{\alpha \beta }= +4 W^{\alpha }{}_{\mu \nu \sigma } \left(f_{30} + f_{31} \Box_s \right) W^{\beta \mu \nu \sigma }$, is given by
\begin{equation}
F_2^{\alpha \beta }=g^{\alpha \beta}\frac{16}{M_{s}^2 \: t^6} \left(
\begin{array}{cccc}
 a_{00}(u) & 0 & 0 & 0 \\
 0 & a_{11}(u) & 0 & 0 \\
 0 & 0 & a_{22}(u) & 0 \\
 0 & 0 & 0 & a_{33}(u) \\
\end{array}
\right)\:,
\end{equation}
with the dimensionless matrix elements $a_{ii}$ defined as
\begin{equation}
a_{00} = a_{11} = a_{22} = a_{33} =+\frac{u^2 (u+1)^2 \left(f_{30} \: t^2 \: M_s^2 \left(u^2+u+1\right)^3  
- 9 f_{31} u^2 (u+1)^2 \right)}{ \left(u^2+u+1\right)^6 }\,. \nonumber
\end{equation}
We can verify, at this point, that terms $F_1^{\alpha\beta}$ and $F_2^{\alpha\beta}$ cancel each other. 

\item The third term,
$F_3^{\alpha \beta }= -4\left(R_{\mu \nu }+\nabla _{\mu }\left.\nabla _{\nu }\right) \left( f_{30} + f_{31} \Box_s \right)W^{\beta \mu \nu \alpha }\right.$, is given by
\begin{equation}
F_3^{\alpha \beta }= g^{\alpha \beta}\frac{16}{M_{s}^2 \: t^6} \left(
\begin{array}{cccc}
 a_{00}(u) & 0 & 0 & 0 \\
 0 & a_{11}(u) & 0 & 0 \\
 0 & 0 & a_{22}(u) & 0 \\
 0 & 0 & 0 & a_{33}(u) \\
\end{array}
\right)\:,
\end{equation}
with the dimensionless matrix elements $a_{ii}$ defined as
\begin{align}
a_{00} =& \frac{ f_{31} \: u^2 (u+1)^2 \left(8 u^6+24 u^5+21 u^4+2 u^3+21 u^2+24 u+8\right)}{\left(u^2+u+1\right)^6} \;,\nonumber \\
a_{11} =& \frac{ f_{31} \: u^2 (u+1)^2 \left(16 u^6+56 u^5+73 u^4+106 u^3+73 u^2+56 u+16\right)}{\left(u^2+u+1\right)^6}\;, \nonumber \\
a_{22} =& \frac{ f_{31} \: u^2 (u+1)^2 \left(16 u^6+40 u^5+33 u^4-54 u^3-127 u^2-120 u-40\right)}{\left(u^2+u+1\right)^6}\;, \nonumber \\
a_{33} =& - \frac{ f_{31} \: u^2 (u+1)^2 \left(40 u^6+120 u^5+127 u^4+54 u^3-33 u^2-40 u-16\right)}{\left(u^2+u+1\right)^6}\;, \nonumber
\end{align}
which only depends on the $f_{31}$ coefficient.

\item The fourth term, $F_4^{\alpha \beta }=  -2f_{31}\nabla ^{\alpha }W^{\lambda }{}_{\mu \nu \sigma }\nabla ^{\beta }W_{\lambda }^{\:\: \mu \nu \sigma } $, is given by
\begin{equation}
F_4^{\alpha \beta }= g^{\alpha \beta}\frac{32}{M_{s}^2 \: t^6} \left(
\begin{array}{cccc}
 a_{00}(u) & 0 & 0 & 0 \\
 0 & a_{11}(u) & 0 & 0 \\
 0 & 0 & a_{22}(u) & 0 \\
 0 & 0 & 0 & a_{33}(u) \\
\end{array}
\right)\:,
\end{equation}
with the dimensionless matrix elements $a_{ii}$ defined as
\begin{align}
a_{00} =& \frac{4 f_{31}   u^2 (u+1)^2  }{\left(u^2+u+1\right)^3}
&& a_{11} = \frac{ f_{31}   u^4 (u^2-1)^2  }{\left(u^2+u+1\right)^6} \nonumber \\
a_{22} =& \frac{ f_{31}   u^2 (u+1)^4 (u+2)^2  }{\left(u^2+u+1\right)^6}
&& a_{33} = \frac{ f_{31}    u^4 (u+1)^4 (2u+1)^2  }{\left(u^2+u+1\right)^6}\,. \nonumber
\end{align}

\item The fifth term, $F_5^{\alpha \beta }=  +g^{\alpha\beta}f_{31}(\n^{\gamma}W_{\lambda}^{\:\:\mu\nu\rho}\n_{\gamma}W^{\lambda}_{\:\:\mu\nu\rho}+W^{\mu\nu\rho\ga}\Box_s W_{\mu\nu\rho\ga}) $, is given by
\begin{equation}
F_5^{\alpha \beta }={\cob -} g^{\alpha \beta}\frac{128}{M_{s}^2 \: t^6} \left(
\begin{array}{cccc}
 a_{00}(u) & 0 & 0 & 0 \\
 0 & a_{11}(u) & 0 & 0 \\
 0 & 0 & a_{22}(u) & 0 \\
 0 & 0 & 0 & a_{33}(u) \\
\end{array}
\right)\:, 
\end{equation}
with the dimensionless matrix elements $a_{ii}$ defined as
\begin{equation}
a_{00} = a_{11} = a_{22} = a_{33} = \frac{ f_{31} \:   u^2 (u+1)^2  }{\left(u^2+u+1\right)^3} \nonumber
\end{equation}

\item The sixth term, $F_6^{\alpha \beta }= -8f_{31}(W^{\ga\nu}{}_{\rho\mu}\n^{\al}W_{\ga}{}^{\bt\rho\mu}-W_{\ga}{}^{\bt\rho\mu}\n^{\al}W^{\ga\nu}{}_{\rho\mu})_{;\nu}  $, is given by
\begin{equation}
F_6^{\alpha \beta }= g^{\alpha \beta}\frac{32}{M_{s}^2 \: t^6} \left(
\begin{array}{cccc}
 a_{00}(u) & 0 & 0 & 0 \\
 0 & a_{11}(u) & 0 & 0 \\
 0 & 0 & a_{22}(u) & 0 \\
 0 & 0 & 0 & a_{33}(u) \\
\end{array}
\right)\:, 
\end{equation}
with the dimensionless matrix elements $a_{ii}$ defined as
\begin{align}
a_{00} =& \frac{ f_{31} \: u^2 (u+1)^2 \left(4 u^6+12 u^5+15 u^4+10 u^3+15 u^2+12 u+4 \right)}{\left(u^2+u+1\right)^6}\;, \nonumber \\
a_{11} =& \frac{ f_{31} \: u^3 ( u^2 - 1)^2 (4 u^2+3 u+4) }{\left(u^2+u+1\right)^6}\;, \nonumber \\
a_{22} =& - \frac{ f_{31} \: u^2 (u+1)^3 (u + 2)^2 (4 u^2+5 u+5) }{\left(u^2+u+1\right)^6}\;, \nonumber \\
a_{33} =& - \frac{ f_{31} \: u^3 (u+1)^3 (2 u +1)^2 (5 u^2+5 u+4) }{\left(u^2+u+1\right)^6}\;. \nonumber
\end{align}
\end{enumerate}
Having computed each term of  $P^{\alpha\beta}_{3}$, up to $\Box_s$, we see that the dependence on 
the $f_{30}$ coefficient vanishes as expected, and the one box, $\Box_s$, contributions survive. 
Finally we have,
\begin{equation}
{P_\Box}_{s3}^{\alpha \beta }= g^{\alpha \beta}\frac{2}{\pi G M_{s}^4 \: t^6} \left(
\begin{array}{cccc}
 a_{00}(u) & 0 & 0 & 0 \\
 0 & a_{11}(u) & 0 & 0 \\
 0 & 0 & a_{22}(u) & 0 \\
 0 & 0 & 0 & a_{33}(u) \\
\end{array}
\right)\:, 
\end{equation}
with the dimensionless matrix elements $a_{ii}$ defined as
\begin{align}
a_{00} =& \frac{ f_{31} \: u^2 (u+1)^2 \left( 16 u^6+48 u^5+51 u^4+22 u^3+51 u^2+48 u+16 \right)}{\left(u^2+u+1\right)^6}\;, \nonumber \\
a_{11} =& \frac{ f_{31} \: u^2 (u+1)^2 \left( 8 u^6+40 u^5+17 u^4+50 u^3+17 u^2+40 u+8 \right)}{\left(u^2+u+1\right)^6}\;, \nonumber \\
a_{22} =& \frac{ f_{31} \: u^2 (u+1)^2 \left( 8 u^6+8 u^5-63 u^4-222 u^3-311 u^2-240 u-80 \right)}{\left(u^2+u+1\right)^6}\;, \nonumber \\
a_{33} =& - \frac{ f_{31} \: u^2 (u+1)^2 \left( 80 u^6+240 u^5+311 u^4+222 u^3+63 u^2-8 u-8 \right)}{\left(u^2+u+1\right)^6}\;. \nonumber
\end{align}
In order to have $P^{\alpha\beta}_{3} = 0$, we should have $a_{00}=a_{11}=a_{22}=a_{33}=0$ for a unique $u \geqslant 1$.
One can show explicitly that there are no any common roots for corresponding algebraic equations.
Therefore, the tensor $P^{\alpha\beta}_{3}$ at one box order can never be made zero by choosing $u$. 

To present even more convincing arguments we put below the answer for second order in box, i.e., $\Box_s^2$ contribution: 
\begin{equation}
	P^{\alpha\beta}_3(\Box_s^2)= g^{\alpha \beta}\frac{24}{\pi G M_{s}^6 \: t^8} \left(
\begin{array}{cccc}
 a_{00}(u) & 0 & 0 & 0 \\
 0 & a_{11}(u) & 0 & 0 \\
 0 & 0 & a_{22}(u) & 0 \\
 0 & 0 & 0 & a_{33}(u) \\
\end{array}
\right)\:, 
\end{equation}
with the dimensionless matrix elements $a_{ii}$ defined as
\begin{align}
a_{00} =& - \frac{ f_{32} \: u^2 (u+1)^2 \left( 16 u^6+48 u^5+53 u^4+26 u^3+53 u^2+48 u+16 \right)}{\left(u^2+u+1\right)^6}\;, \nonumber \\
a_{11} =& - \frac{ f_{32} \: u^2 (u+1)^2 \left( 48 u^8+232 u^7+457 u^6+735 u^5+812 u^4+735 u^3+457 u^2+232 u+48 \right)}{\left(u^2+u+1\right)^7}\;, \nonumber \\
a_{22} =& - \frac{ f_{32} \: u^2 (u+1)^2 \left( 48 u^8+152 u^7+177 u^6-177 u^5-768 u^4-1129 u^3-899 u^2-448 u-112 \right)}{\left(u^2+u+1\right)^7}\;, \nonumber \\
a_{33} =& \frac{ f_{32} \: u^2 (u+1)^2 \left( 112 u^8+448 u^7+899 u^6+1129 u^5+768 u^4+177 u^3-177 u^2-152 u-48 \right)}{\left(u^2+u+1\right)^7}\;. \nonumber
\end{align}
As in the case of the first order in $\Box_s$, there are no common roots for $u$, which would allow all the above polynomials to vanish.

\acknowledgments 
AK and JM are supported by the grant UID/MAT/00212/2013 and COST Action CA15117 (CANTATA). AK is supported by FCT Portugal investigator project IF/01607/2015 and FCT Portugal fellowship SFRH/BPD/105212/2014.



\end{document}